\documentclass[useAMS,usenatbib]{mn2e}
\usepackage{aas_macros}
\usepackage[a4paper,centering, totalwidth=520pt, totalheight=700pt]{geometry}

\usepackage{graphicx}
\usepackage{amssymb}
\usepackage{amsmath}
\usepackage{enumerate}
\usepackage{microtype}

\usepackage{color}

\newcommand{\del}{\ensuremath{\delta}}
\newcommand{\Del}{\ensuremath{\Delta}}

\newcommand{\gam}{\ensuremath{\gamma}}

\newcommand{\sig}{\ensuremath{\sigma}}

\newcommand{\epc}{\ensuremath{\epsilon_{\times}}}
\newcommand{\Sc}{\ensuremath{S_{\times}}}
\newcommand{\So}{\ensuremath{S_{0}}}
\newcommand{\delc}{\ensuremath{\delta_{\rm c}}}
\newcommand{\delo}{\ensuremath{\delta_{0}}}

\newcommand{\fcoll}{\ensuremath{f_{\rm coll}}}
\newcommand{\HII}{\ensuremath{H_{\rm II}}}

\newcommand{\avg}[1]{\ensuremath{\left\langle \,#1\, \right\rangle}}

\newcommand{\der}{\ensuremath{{\rm d}}}

\newcommand{\erfc}[1]{\ensuremath{{\rm erfc}\left(#1\right)}}
\newcommand{\erf}[1]{\ensuremath{{\rm erf}\left(#1\right)}}

\newcommand{\eqn}[1]{equation~\eqref{#1}}

\newcommand{\ph}[1]{\phantom{#1}}

\newcommand{\be}{\begin{equation}}
\newcommand{\ee}{\end{equation}}

\newcommand{\Cal}[1]{\ensuremath{\mathcal{#1}}}

\begin{document}

\title[Modelling HII bubbles during reionization]
{An improved model of {\sc Hii} bubbles during the epoch of reionization}
\author[Paranjape \& Choudhury]
{Aseem Paranjape$^1$\thanks{Email: aseemp@phys.ethz.ch}~
and
T. Roy Choudhury$^2$\thanks{E-mail: tirth@ncra.tifr.res.in}~
\\
$^1$ETH Zurich, Department of Physics, Institute for Astronomy, Wolfgang-Pauli-Strasse 27, CH-8093 Zurich, Switzerland\\
$^2$National Centre for Radio Astrophysics, TIFR, Post Bag 3, Ganeshkhind, Pune 411007, India
} 

\maketitle

\date{\today}

\begin{abstract}
The size distribution of ionized regions during the epoch of reionization --  a key ingredient in understanding the {\sc Hi} power spectrum observable by 21cm experiments -- can be modelled analytically using the excursion set formalism of random walks in the smoothed initial density field. To date, such calculations have been based on simplifying assumptions carried forward from the earliest excursion set models of two decades ago. In particular, these models assume that the random walks have uncorrelated steps and that haloes can form at arbitrary locations in the initial density field. We extend these calculations by incorporating recent technical developments that allow us to (a) include the effect of correlations in the steps of the walks induced by a realistic smoothing filter and (b) more importantly, account for the fact that dark matter haloes preferentially form near \emph{peaks} in the initial density. A comparison with previous calculations shows that including these features, particularly the peaks constraint on halo locations, has large effects on the size distribution of the {\sc Hii} bubbles surrounding these haloes. For example, when comparing models at the same value of the globally averaged ionized volume fraction, the typical bubble sizes predicted by our model are more than a factor 2 larger than earlier calculations. 
Our results can potentially have a significant impact on estimates of the observable {\sc Hi} power spectrum.
\end{abstract}


\section{Introduction}
\label{outline}
\noindent
An outstanding problem in present day cosmology is to understand the reionization of neutral hydrogen that occurs between $20 \gtrsim z \gtrsim 6$ \citep{2006PhR...433..181F,pl12}. From the observational viewpoint, the study of the reionization epoch is extremely interesting because this is the last phase of cosmic evolution that can be observed directly. Theoretically, reionization is linked to the formation of the first galaxies and stars. It is the ionizing photons from these sources which are believed to be the drivers of the initial phase of reionization; observations of the reionization history could therefore yield important information about the nature and evolution of these sources. These ionizing sources are expected to form ionized regions (bubbles, in what follows) around themselves, which then overlap and proceed to complete the process of reionization. According to current models, this process should have occurred over the redshift range $20 \gtrsim z \gtrsim 6$ \citep{2003ApJ...586..693W,2005MNRAS.361..577C,2006MNRAS.371L..55C,2006ApJ...644L.101A,2010MNRAS.408...57P,2011MNRAS.413.1569M,2012MNRAS.419.1480M}.

Observationally, constraining reionization at $z \gtrsim 6$ has been extremely challenging for various reasons. For example, the mean transmittance of Lyman-$\alpha$ flux in the quasar absorption spectra becomes so low \citep{2006AJ....132..117F} that the constraints on the neutral hydrogen fraction and photoionization rate turn out to be highly model-dependent \citep{2006MNRAS.370.1401G,2007MNRAS.382..325B,2008MNRAS.386..359G,2008MNRAS.388L..84G,2011MNRAS.416L..70B}. Similarly, fluctuations in the CMBR polarization at large angular scales are expected to be cosmic variance dominated, and hence it is unlikely that CMBR observations would be able to provide more information on reionization than what is available at present \citep{2003ApJ...595...13H,2012MNRAS.419.1480M}.
A potential way to constrain the reionization history at high redshifts is by detecting the 21cm signal from neutral hydrogen ({\sc Hi}) at those redshifts. A large effort is being directed towards this end using radio interferometric instruments such as GMRT\footnote{http://www.gmrt.ncra.tifr.res.in/} \citep{2011MNRAS.413.1174P,2013MNRAS.433..639P}, LOFAR\footnote{http://www.lofar.org/} \citep{2010MNRAS.405.2492H,2013A&A...550A.136Y}, MWA\footnote{http://www.haystack.mit.edu/ast/arrays/mwa/} \citep{2013PASA...30....7T,2013PASA...30...31B}, etc. In addition, one expects significant breakthroughs from future radio telescopes such as the SKA\footnote{http://www.skatelescope.org/} \citep{2013ExA....36..235M}.

Since the size distribution of the ionized bubbles depends on the nature of the sources from which they form, and in turn determines the observable 21cm signal, building analytical models of the evolution of this distribution can give us a handle on interpreting future observations and constraining astrophysical and cosmological models. This has been the motivation behind much recent analytical 
\citep{fzh04,2004ApJ...608..622Z,2006MNRAS.365..115F,2006ApJ...653..815M,2007MNRAS.375.1034W,2007MNRAS.379.1647W}
and numerical work on this subject \citep{2006MNRAS.369.1625I,2006MNRAS.372..679M,2007ApJ...654...12Z,2007MNRAS.377.1043M,2007ApJ...671....1T,2007A&A...474..365S,2008ApJ...681..756S,2008ApJ...680..962L,2009A&A...495..389B,2009MNRAS.393...32T,2009MNRAS.394..960C}.
Since the physics governing the formation of the first sources is poorly understood, theoretical models often contain a large number of uncertain parameters \citep{2010MNRAS.408...57P,2011MNRAS.413.1569M,2012MNRAS.419.1480M,2012ApJ...747..100S}. In this regard, analytical or semi-analytical models are therefore very useful in probing large regions of parameter space. 

A particular class of models, which estimate the size distribution of these bubbles and their 21cm signals, are based on the excursion set approach \citep{ps74,e83,ph90,bcek91} which has traditionally been applied to the problem of estimating the mass function, growth and clustering of dark matter haloes \citep{lc93,mw96,bm96,Sheth98,Monaco99,smt01,zh06,mr10,pls12,ms12,arsc13}. In this context, this approach essentially maps a counting problem in the evolved dark matter density field to the statistics of random walks in the smoothed \emph{initial} density field as a function of smoothing scale, through some simplifying assumption on the dynamical evolution of gravitational nonlinearities such as spherical \citep{gg72,ps74} or ellipsoidal \citep{bm96,smt01} collapse. 
The latter leads to a density threshold for the random walks, and the mass fraction in haloes of mass $m$ is equated to the `first-crossing distribution', i.e., the fraction of walks that first cross this threshold at Lagrangian smoothing scale $R_{\rm L}\propto m^{1/3}$ as the scale is decreased from large values.
Since the initial density field is expected to be close to Gaussian, this allows for simple approximate solutions for the abundance and clustering of halos.

\citet*[][hereafter, FZH04]{fzh04} first pointed out that, by treating collapsed dark matter haloes as the hosts of the sources of reionization, essentially the same counting problem applies to the growth of ionized {\sc Hii} bubbles during reionization as well, and can be used to estimate the size distribution of these bubbles under some simplifying assumptions. (We describe their model in detail below.) Under the excursion set ansatz, similarly to the case of haloes, the comoving number density $\der n/\der\ln R_0$ of ionized bubbles of Lagrangian volume $V_0=4\pi R_0^3/3$ is mapped to the first-crossing distribution $f$ of a specific threshold or `barrier' by random walks in the smoothed initial density \delo\ as the smoothing scale $R_0$ is decreased from large values: 
\be
V_0 \frac{\der n}{\der\ln R_0} = S_0 f(S_0)\,\left|\frac{\der\ln\So}{\der\ln R_0}\right|\,,
\label{V0dndlnR0}
\ee
where $\So=\avg{\delo^2}$ is the variance of density fluctuations on scale $R_0$.

One set of assumptions inherent in the FZH04 analysis derives from the original excursion set formalism presented by \citet{bcek91} and pertains to the nature of the random walks; these walks are assumed to have steps that are uncorrelated with each other, which would be the case if the smoothing filter for the density were sharp in Fourier space. This introduces some technical simplifications in the analysis, while using a more realistic filter such as a TopHat in real space leads to correlations in the steps which are difficult to treat in full generality. Additionally, the FZH04 calculation explicitly uses the \citet{bcek91} expression for the mass fraction in collapsed objects, which relies on the spherical collapse model and further treats all locations in the initial density field on equal footing, whereas one expects haloes to form preferentially near \emph{peaks} in the initial density \citep{bbks86,Bond1989} with a dynamical evolution that is better described by ellipsoidal collapse \citep{bm96,smt01}.

Recent work has shown that there exist simple yet remarkably accurate approximations that account for the correlations in the steps of the random walks induced by a realistic filter \citep{ms12,ms13}. These approximations make it relatively simple to additionally include the effect of centering the walks at \emph{peaks} in the initial density \citep{ms12,ps12} as well as including the effects of ellipsoidal collapse, leading to a modified approach of Excursion Set Peaks (ESP) that agrees with halo abundances and clustering measurements in $N$-body simulations at the $\sim10\%$ level \citep{psd13}. These improvements, which we describe in detail below, form the basis of the current work\footnote{
Another complication is that the excursion set ansatz formally considers random walks centered at a fixed location for an ensemble of initial conditions and therefore potentially misses the effects of correlations \emph{between} walks centered at different locations. E.g., the numerical algorithm of \cite{bm96} explicitly accounts for these correlations, and there have been attempts to account for these effects analytically for sharp-$k$ filtering \citep[see, e.g.,][]{sb02,Sheth2011}. 
Recent work \citep{hp14} has shown, however, that the ESP framework of \citet{psd13} approximates the effects of these `correlated walks' remarkably well when compared with numerical results similar to those of \citet{bm96}.
Since the ESP approach is a simpler and more realistic solution of this problem, we will not explore alternate formalisms here.
}.  

Our main idea in this paper is to investigate the effects these improvements have on the size distribution of the {\sc Hii} bubbles. The FZH04 model, with its assumptions of sharp-$k$ filtering and spherical collapse at randomly chosen locations, formed the basis for generating semi-numerical ionization maps in a three-dimensional box \citep{2007ApJ...669..663M,2008MNRAS.386.1683G,2011MNRAS.411..955M}, with later improvements in the semi-numerical calculations that included sharp-$k$ filters consistently throughout the calculations \citep[e.g., the FFRT scheme of][]{2011MNRAS.414..727Z}. These semi-numerical estimates of the ionization maps were found to match the results of radiative transfer simulations to some degree of accuracy \citep{2007ApJ...654...12Z,2011MNRAS.414..727Z,2011MNRAS.411..955M}. It is almost a decade since the first simple analytical treatments were introduced, which were useful in building intuition. Since there have been technical improvements in modelling dark matter on the analytical side as mentioned above, we believe it is worth repeating the calculation to see whether the improvements transfer to the bubble distributions too. We focus on analytical results here and leave a detailed comparison with numerical models to future work.

The plan of the paper is as follows. In section~\ref{sec:formalism} we describe the analytical excursion set formalism, starting with a recapitulation of the FZH04 model in section~\ref{subsec:fzhrecap}, followed by a description of the ESP framework and the associated modifications to the bubble size distribution calculation in section~\ref{subsec:peaks}. 
The results of the comparison between the two are described in Section~\ref{sec:results}. We perform two kinds of comparisons, one when assuming the same astrophysical ionization efficiency for both models (section~\ref{subsec:samezeta}) and another where we fix the globally averaged ionized volume fraction to be the same in both models (section~\ref{subsec:sameQ}). We conclude in section~\ref{sec:conclude}.

\section{Formalism}
\label{sec:formalism}
In this section we describe the formalism we use to estimate the size distribution of ionized bubbles. We start with a summary of the approach presented by FZH04, and then describe the modifications introduced by considering peaks theory and random walks with correlated steps.

\subsection{FZH04 calculation with sharp-$k$ walks}
\label{subsec:fzhrecap}
The basic assumptions that go into the FZH04 calculation are that $(i)$ any halo
above a minimum mass $m_{\rm min}$ is an ionizing source and $(ii)$
such a halo of mass $m$ ionizes a mass $m_{\rm ion} = \zeta m$ with
$\zeta>1$ an efficiency factor related to various astrophysical
quantities. In the following we will frequently use the relation between mass scale $m$, Lagrangian radius $R_{\rm L}$ and the variance $\sig_0^2$ of the smoothed linearly extrapolated density contrast given by
\be
\sig^2_0(m) \equiv \avg{\del_{R_{\rm L}}^2} = \int\der\ln k~\Del^2(k)W^2(kR_{\rm L})\,,
\label{sig0-def}
\ee
where $\Del^2(k)\equiv k^3P(k)/2\pi^2$ is the dimensionless matter power spectrum linearly extrapolated to present epoch, $W(kR_{\rm L})$ is the smoothing filter in Fourier space, and the Lagrangian radius is related to the mass through
\be
m = (4\pi/3)\bar\rho R_{\rm L}^3\,,
\label{m-RL}
\ee
with $\bar\rho$ the present mean density of the Universe.

Let $f(m|M,V)\der m$ be the mass fraction in haloes having mass within $m$ and $m + \der m$ embedded in a region of Eulerian volume $V$ which encloses a total mass $M=(4\pi/3)\bar\rho R_0^3$. Then, denoting $\So=\sig_0^2(M)$ and $s=\sig_0^2(m)$, we can change variables to write
\be
f(m|M,V)\der m = f(s|\delo,\So)\der s\,. 
\label{confc-mass2var}
\ee
The quantity $f(s|\delo,\So)$ is typically calculated using a conditional first crossing distribution of a chosen barrier, by random walks in the density contrast that are constrained to pass through \delo\ on scale \So.

In writing \eqn{confc-mass2var}, we have implicitly assumed that there is a one-to-one mapping between Lagrangian regions of scale \So\ that have a linear density contrast \delo, and Eulerian regions of (comoving) volume $V=4\pi R^3/3$ that enclose mass $M$ (and therefore have a nonlinear density contrast $\del_{\rm NL}=M/(\bar\rho V) - 1$). This is true, e.g., if one assumes spherical evolution of the volume \citep{Sheth98}. In this case, assuming an Einstein-deSitter background (which is an excellent approximation at the redshifts of interest), the comoving Lagrangian radius $R_0$ of the sphere is related to its physical Eulerian radius $R_{\rm phys}$ at redshift $z$ through \citep{gg72}
\be
\frac{R_{\rm phys}}{R_0} = \frac{3}{10\delo}\left(1-\cos\theta\right);~
\frac1{1+z} = \frac{3\cdot6^{2/3}}{20\delo}\left(\theta-\sin\theta\right)^{2/3},
\label{sphcoll}
\ee
and its comoving Eulerian radius is $R = (1+z)R_{\rm phys}$. We have assumed $\delo>0$ since we will not consider situations in which underdense voids contain enough sources to be fully ionized. By the time this occurs, the ionized regions are expected to have percolated through most of the volume in the Universe, with reionization progressing very rapidly. It is not clear whether the simplifying assumptions of the excursion set calculation remain valid in this regime. We will return to this issue in future work where we compare our analytical models with numerical calculations.

The mass fraction contained in all ionizing sources inside the volume $V$ is 
\begin{align}
\fcoll(M,V) &\equiv \int_{m_{\rm min}}^M\der m\,f(m|M,V) \notag\\
&=\int_{\So}^{s_{\rm min}}\der s\,f(s|\delo,\So) 
= \fcoll(\delo,\So)\,,
\label{fcoll}
\end{align}
where $s_{\rm min}\equiv \sig_0^2(m_{\rm min})$. The mass fraction \emph{ionized} by these sources is therefore $\zeta\,\fcoll$, which means that a region can be treated as completely ionized when
\be
\zeta\,\fcoll(M,V) \geq 1\,.
\label{fullyionized}
\ee
This, however, does not account for the effect of ``extra'' ionizing photons from neighbouring regions. The idea here is that if a region has a collapse fraction such that $\zeta\fcoll\gg1$, then some of the ionizing photons can ``leak'' into neighbouring regions.
To include this effect, FZH04 proposed that an ionized region be assigned a volume $V$ provided that $V$ is the \emph{largest} volume for which the condition \eqref{fullyionized} is satisfied, which prevents overcounting of overlapping/nested ionized bubbles\footnote{Of course this is only approximate since real bubbles are not expected to be spherical. Additionally, the assumption of a deterministic value for $\zeta$ is also known to be an oversimplification. Since our goal in this paper is only to compare relative differences between two analytical models, we will retain these simplifying assumptions and leave a more realistic comparison with numerical results to future work.}. This counting problem is therefore very similar to the one of counting Lagrangian regions that will eventually form halos, and can be treated in the same excursion set language. There are some differences between the two counting problems, however, which we will discuss below.

The FZH04 calculation is based on the original treatment by \citet{bcek91} which studies random walks in the overdensity \del\ smoothed using a filter that is sharp in $k$-space\footnote{Note that in usual applications of the \citet{bcek91} formalism one assumes a sharp-$k$ filter when calculating the mass fraction but a real-space TopHat filter when calculating the variance $\sig_0^2(m)$. The same assumption is implied in the original FZH04 calculations. In our later calculation we will self-consistently use real-space TopHat filtering for both \del\ and $\sig_0^2(m)$.}. Further assuming the mass-independent critical overdensity or ``barrier'' $\delc(z)$ predicted by spherical collapse, this leads to a closed form expression for the mass fraction given by
\begin{align}
f_{\rm coll,sh-k}(M,V) &= \int_{\So}^{s_{\rm min}}\der
s\,f(\delc(z)-\delo;s-\So) \notag\\
&=\erfc{\frac{\delc(z)-\delo}{\sqrt{2(s_{\rm min}-\So)}}}\,,
\label{fcoll-sharpk}
\end{align}
where $f(\delc(z),s(m))$ gives the uncondtional mass fraction in halos of mass $m$ at redshift $z$, and the first equality follows from the fact that sharp-$k$ filtered walks have uncorrelated steps, so that the conditional mass fraction is simply the unconditional one evaluated with a simple ``shift of origin'' \citep{lc93}. This exact result will no longer be true when considering realistic filters below.

An implicit assumption here is that at very large scales where $\So\to0$ and $\delo\to0$, one ends up with the condition $\erfc{\delc(z)/\sqrt{2s_{\rm min}}} < \zeta^{-1}$. Guaranteeing this implies a restriction on allowed $(\zeta,m_{\rm min})$ values. A similar restriction will also apply in our later calculation including the peaks constraint. Since the error function is a monotonic function of its argument, the condition \eqref{fullyionized} becomes an inequality on \delo,
\be
\delo \geq \delc(z) - \sqrt{2}K(\zeta)\left(s_{\rm min}-\So\right)^{1/2}
\equiv B_{\rm FZH}(\So;\zeta,z)\,,
\label{fc-ineq-FZH}
\ee
where $K(\zeta) = {\rm erfc}^{-1}(\zeta^{-1}) = {\rm erf}^{-1}(1-\zeta^{-1})$. Hence the problem of locating ionized bubbles is reduced to solving a first passage problem with the moving barrier $B_{\rm FZH}(\So;\zeta,z)$. FZH04 did this by using the same sharp-$k$ formalism applied to a linear approximation to this barrier, 
\be
B_{\rm FZHlin}(\So;\zeta,z) = B_0 + B_1\,\So
\label{B-FZH-lin}
\ee
where $B_0=\delc(z)-K(\zeta)\sqrt{2s_{\rm min}}$ and $B_1=K(\zeta)/\sqrt{2s_{\rm min}}$. With this approximation, the solution for the first crossing distribution is given by \citep{Sheth98}
\be
f_{\rm FZH}(\So;\zeta) = \frac{B_0}{\sqrt{2\pi\So^3}}{\rm e}^{-B_0^2/2\So}{\rm e}^{-B_0B_1}{\rm e}^{-B_1^2\So/2}\,.
\label{fc-FZH-lin}
\ee
The Lagrangian bubble size distribution then follows from setting $f(\So)\to f_{\rm FZH}(\So;\zeta)$ in \eqn{V0dndlnR0}.

A useful quantity is the fraction of Lagrangian volume (i.e., the mass fraction) filled by the ionized bubbles; in this model this is given by
\begin{align}
Q_0 &= \int_{R_{\rm 0,min}}^\infty\der\ln R_0\left(V_0\frac{\der n}{\der\ln R_0}\right)\notag\\
  &= \frac{1}{2}\left[{\rm e}^{-2b}\erfc{\frac{\nu^2_{\rm 0,min}-b}{\sqrt{2}~\nu_{\rm 0,min}}} + \erfc{\frac{\nu^2_{\rm 0,min} + b}{\sqrt{2}~\nu_{\rm 0,min}}}\right]\,,
\label{QLag-FZH}
\end{align}
where $R_{\rm 0,min}=(3\zeta m_{\rm min}/4\pi\bar\rho)^{1/3}$ is the smallest possible Lagrangian bubble size, and in the second line we defined $b=B_0B_1$ and $\nu_{\rm 0,min}=B_0/\sig_0(\zeta m_{\rm min})$. Notice that, contrary to what would be expected based on photon number conservation arguments \citep{fzh04}, in general this expression for $Q_0$ is not equal to $\zeta\erfc{\delc(z)/\sqrt{2s_{\rm min}}}$. This is a generic feature of excursion set calculations and is shared by the peaks-based approach we describe below. Although this issue of non-conservation of photon number has been discussed in the past \citep[e.g.,][]{2007ApJ...654...12Z}, we believe it deserves a more thorough investigation which we leave to future work.

To summarize, adopting the traditional sharp-$k$ excursion set analysis has the following effects. Firstly, this fixes the conditional mass fraction \fcoll\ as a complementary error function, and the form of its argument in terms of \delo\ and \So. Next, this fixes the shape of the barrier $B_{\rm FZH}(\So;\zeta)$. And finally, it fixes the solution to the first passage problem in the presence of this barrier. 

\subsection{Accounting for peaks and correlated steps}
\label{subsec:peaks}
The original excursion set formulation of \citet{bcek91} has two shortcomings. Firstly, it uses sharp-$k$ filtering when studying random walks in \del, which leads to some technical simplifications (the walks have uncorrelated rather than correlated steps) but is less realistic than using real-space TopHat filtering. Secondly, the traditional excursion set formalism is based on counting randomly placed regions, while halos preferentially form near initial density peaks: this has been the guiding principle behind several works in the past \citep{bbks86,aj90,bm96,m+98,h01} and has recently been explicitly verified in simulations \citep{lp11,hp14}. Ignoring the fact that the locations of collapse are special and not arbitrarily placed is known to lead to systematically wrong predictions for the final halo mass function \citep{smt01}. To improve upon this calculation, we must therefore account for these two effects: both first passage problems (for halos as well as ionized regions) must now be solved with correlated steps due to a nontrivial filter $W(kR_{\rm L})$, and the conditional mass fraction \fcoll\ must be calculated with the additional constraint that the walks describing halos be centered on peaks\footnote{The ionized regions needn't be centered on density peaks, so the peaks constraint will not enter there.}. 

Recent work \citep[][MS12]{ms12} has shown that the effects of correlations arising from realistic filtering such as the TopHat or Gaussian can be accurately described using a simple approximation. Essentially, this involves recognizing that the filter induces \emph{strong} correlations between the steps of the walks \citep{pls12}. This means that the first-crossing condition that solves the cloud-in-cloud problem and, in principle, corresponds to an infinite number of constraints, can be replaced with the simpler \emph{up}-crossing condition \citep{Bond1989,bcek91} which requires exactly two constraints and can be solved analytically \citep[see also][]{ms13}. This also allows the peaks constraint \citep{bbks86} to be included in a straightforward way \citep[MS12;][]{ps12} and leads to an analytical prescription (Excursion Set Peaks; ESP) for the halo mass function and halo bias which describes the results of $N$-body simulations to within $\sim10\%$ \citep{psd13}. 

In Appendix~\ref{app:esp} we describe these calculations in some detail. The conditional mass fraction in the ESP case works out to 
\be
\fcoll(M,V) = \int_{\So}^{s_{\rm min}}\der s\,f_{\rm ESP}(s|\delo,\So)\,.
\label{fcoll-ESP}
\ee 
where $f_{\rm ESP}(s|\delo,\So)$ is given by \eqn{fESPcond}. 
The condition for identifying ionized regions \eqref{fullyionized} can be now generalized to a condition 
\be
\delo \geq B_{\HII}(\So;\zeta,z)\,.
\label{fi-d0}
\ee
For each \So, the barrier $B_{\HII}(\So;\zeta,z)$ is the value of \delo\ where $\zeta\fcoll(\delo,\So)=1$, which must be solved for numerically since there is no closed form expression for the conditional mass fraction $f_{\rm ESP}(s|\delo,\So)$. This is the most time-consuming part of the calculation.

Knowing the barrier $B_{\HII}(\So;\zeta,z)$, the first passage problem \eqref{fi-d0} can be solved using the MS12 approximation: 
\be
f_{\HII}(\So;\zeta) = \int_{B_{\HII}^\prime}^\infty\der
v\,(v-B_{\HII}^\prime) p(B_{\HII},v;\So)\,, 
\label{fc-HII}
\ee
where $v\equiv\der\delo/\der\So$, $B_{\HII}^\prime\equiv\der B_{\HII}/\der\So$ and $p(B_{\HII},v;\So)$ is a bivariate Gaussian with a covariance matrix fixed by the correlation between the density contrast $\delo$ and its derivative $v$ at scale \So. This expression can be written in closed form (equation 5 of MS12) and, in the present case, is valid for $0 < \So < \sig_0^2(\zeta m_{\rm min})$. 

The approximations inherent in the MS12 (and ESP) analytical formulae will break down if the value of \delo\ approaches that of the halo
barrier \citep[see Figure 3 of][]{mps12}. This will never happen if $B_{\HII}(\So;\zeta,z) < B_{\rm halo}(\So;z)$ which is the case here.

\subsection{From Lagrangian to Eulerian coordinates}
So far we have not explicitly discussed how to convert the Lagrangian result \eqref{fc-HII} into the distribution of Eulerian bubble sizes implied by our notation $\fcoll(M,V)$. 
Making the relation explicit requires an extra step, namely, we assume the regions to be spherical and then use equation \eqref{sphcoll} replacing \delo\ with the value of the barrier $B_{\HII}(\So;\zeta,z)$.

The Lagrangian calculation gives us a comoving number density $\der n/\der\ln R_0$ of bubbles of size $R_0$ or volume $V_0$ given by \eqn{V0dndlnR0} with $f(\So)\to f_{\HII}(S_0;\zeta)$.
The Lagrangian to Eulerian mapping preserves the comoving number density of the bubbles, since the comoving volume of the Universe (or simulation box) remains constant and there is a one-to-one mapping between the Lagrangian and Eulerian radius of each bubble. So we have
\be
V \frac{\der n}{\der\ln R} = \frac{V}{V_0} \left(\frac{\der\ln R}{\der\ln R_0}\right)^{-1} S_0 f_{\HII}(S_0;\zeta)\,\left|\frac{\der\ln\So}{\der\ln R_0}\right|\,.
\label{VdndlnR}
\ee
The Jacobian $\der\ln\So/\der\ln R_0$ is standard, and the ratio $V/V_0$ follows from \eqn{sphcoll}. All we need then is the Jacobian $\der\ln R/\der\ln R_0$ which is given by
\be
\frac{\der\ln R}{\der\ln R_0} = 1 - \left|\frac{\der\ln\So}{\der\ln R_0}\right|\frac{\der\ln B_{\HII}}{\der\ln \So}\left(1-\frac32\frac{\theta(\theta-\sin\theta)}{(1-\cos\theta)^2}\right)\,.
\label{dlnRdlnR0}
\ee
Converting from Lagrangian to Eulerian coordinates can potentially lead to quantitative differences in the calculations at the end since bubbles are defined by overdense regions and hence their expansion could be different from the background.

As with the FZH04 calculation, the Lagrangian volume fraction in ionised bubbles is given by $Q_0=\int_{R_{\rm 0,min}}^\infty\der\ln R_0 \,(V_0\der n/\der\ln R_0)$ with $R_{\rm 0,min}=(3\zeta m_{\rm min}/4\pi\bar\rho)^{1/3}$; this needs a numerical calculation in the ESP case. The corresponding Eulerian volume fraction is
\begin{align}
Q &= \int_{R_{\rm min}}^{\infty} \der \ln R\left(V\frac{\der n}{\der \ln R}\right)\notag\\
  &= \int_{R_{\rm 0,min}}^\infty\der\ln R_0\left(V_0\frac{\der n}{\der\ln R_0}\right)\left(\frac{V}{V_0}\right)
\label{Q-def}
\end{align}
where $R_{\rm min}$ is the Eulerian size corresponding to $R_{\rm 0,min}$. Below we will study the normalized bubble size distributions $Q_0^{-1} V_0 \der n/\der \ln R_0$ and $Q^{-1} V \der n/\der \ln R$.

\section{Results}
\label{sec:results}
\noindent
We present the main results of our calculation in this section. 
We will consider Gaussian initial conditions and a flat Lambda-cold dark matter (LCDM) cosmology throughout, with
three different sets of parameters which are summarized in Table \ref{tab:cosmo}.
The parameters $\Omega_{\rm m}$ and $\Omega_{\rm b}$ are the present total and baryonic matter fractions, respectively, the Hubble constant is $H_0=100h\,$km/s/Mpc, $\sigma_8$ is the r.m.s. of linear density fluctuations smoothed in spheres of size $8h^{-1}$Mpc, and $n_{\rm s}$ is the scalar spectral index.
Our fiducial cosmology, which we refer to as Planck13, is compatible with the recent results from the \citet{Planck13}. Additionally, we will consider two other cosmologies, one we refer to as WMAP1 \citep{2003ApJS..148..175S} and the other as WMAP3 \citep{2007ApJS..170..377S}. 
One can see from the Table that the primary difference between Planck13 and WMAP1 is in the value of $\sig_8$ while WMAP3 and Planck13 differ mainly in the values of $\Omega_{\rm m}$ and $\Omega_{\rm b}$. 
(The WMAP1 parameter values are the same as those used by FZH04.)
We use {\sc Camb} \citep*{camb}\footnote{   http://lambda.gsfc.nasa.gov/toolbox/tb\_camb\_form.cfm} to generate all transfer functions. The value of $m_{\rm min}$, which appears in expressions for the collapsed fraction, is chosen such that only atomically cooled haloes are allowed to form stars (we set the associated virial temperature to $T_{\rm vir}=10^4$K.). 

Now, assuming we fix the cosmology to one of the above, it is possible to compare bubble size distributions at a given $z$ in two different ways: (i) we can fix the value of the efficiency parameter $\zeta$ which will give rise to different values of $Q$ at the given redshift, or (ii) we can normalize the curves to the same value of $Q$ by using a different value of $\zeta$ and then compare. The first approach is conceptually straightforward as it captures the difference in size distributions when the astrophysical properties of the sources remain the same. The advantage of the second method
is that it normalizes all models to the same value of the global ionization
fraction, and hence the differences would be only in the topology of the
ionized regions. We will present results using both the approaches, first when $\zeta$ is kept constant, and next with $Q$ kept fixed.

\begin{table}
\centering
\begin{tabular}{cccccc}
\hline
Name & $\Omega_{\rm m}$ & $\sigma_8$ & $n_{\rm s}$ & $h$ & $\Omega_{\rm b}$\\
\hline
Planck13 & 0.315 & 0.829 & 0.960 & 0.673 & 0.0487 \\
WMAP1    & 0.30 & 0.90 & 1.00 & 0.70 & 0.046 \\
WMAP3    & 0.25 & 0.80 & 1.00 & 0.70 & 0.040 \\
\hline
\end{tabular}
\caption{Cosmological models and parameters used in the paper.}
\label{tab:cosmo}
\end{table}

\subsection{Models with the same $\zeta$}
\label{subsec:samezeta}
\noindent
In this section, we compare the predictions of different models for 
the same value of $\zeta$, which will be constant at $\zeta=40$ in our fiducial model.
We first compare the barriers, ionized volume fractions and bubble size distributions predicted by the ESP and FZH04 models, and then study the effect of varying the cosmological parameters and form of $\zeta$ within the ESP model.

\subsubsection{Comparison of barriers}
\label{subsec:barriers}
\begin{figure}
\centering
\includegraphics[width=0.45\textwidth]{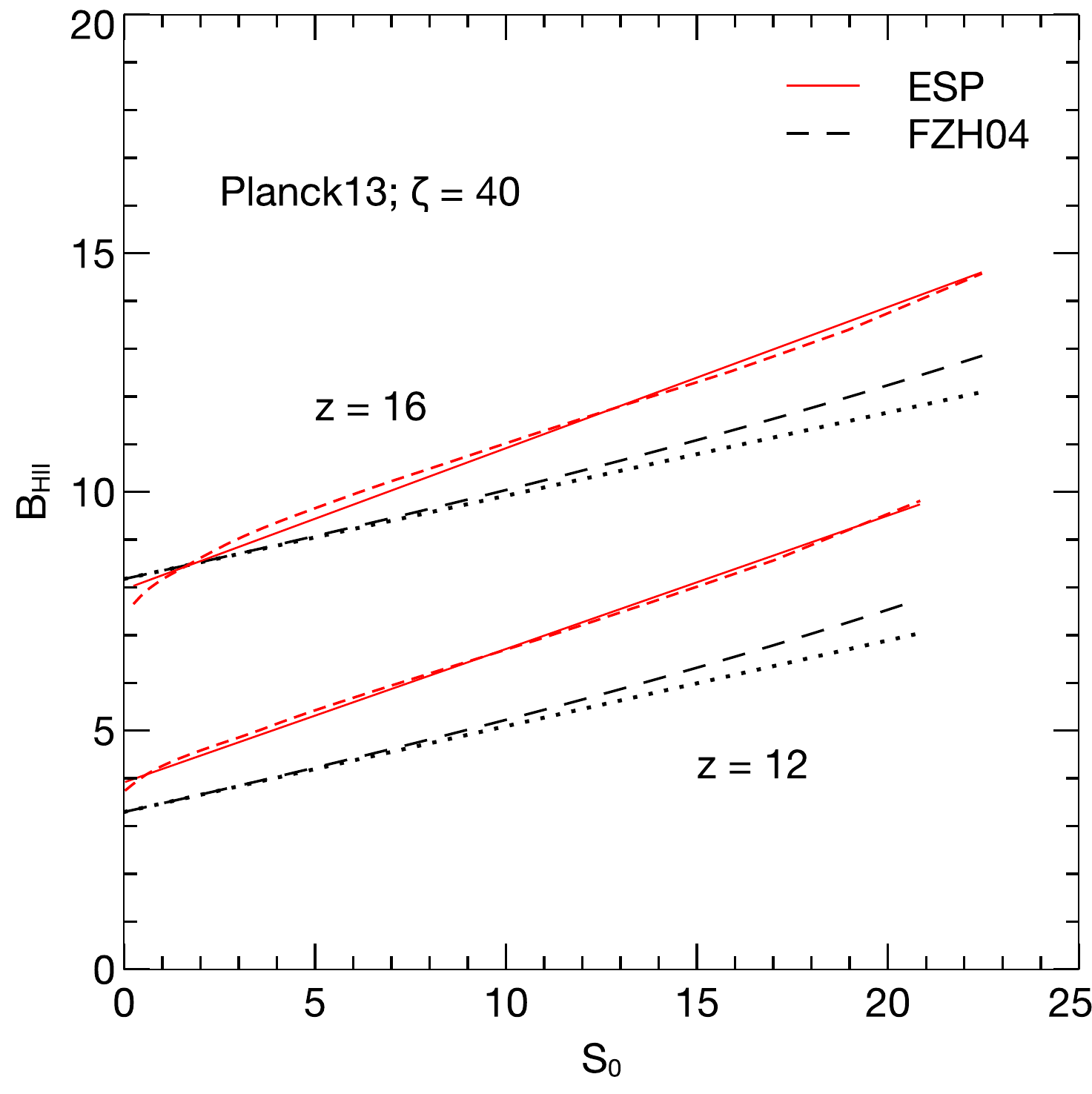}
\caption{The ionization barrier for the FZH04 (dashed black) and ESP (fine-dashed red) cases for two representative
redshifts $z=12, 16$. The cosmology chosen is Planck13 and the value of $\zeta = 40$. Also shown are the linear approximations to the barriers (dotted black for FZH04 and solid red for ESP). See text for details.}
\label{fig:barriers}
\end{figure}

\noindent
Figure~\ref{fig:barriers} shows the behaviour of the barrier $B_{\rm HII}$ for different cases.
As we will see later, many of the results can be understood from the 
heights and shapes of these barriers.
We show the {\sc Hii} barrier for the FZH04 (dashed black) and ESP (fine-dashed red) models for two
different redshifts $z=12,16$ for $\zeta = 40$. 
Physically, ionized bubbles form hierarchically with smaller bubbles forming first and then growing and merging to form larger bubbles. The excursion set approach very naturally accounts for this feature, which is very similar (and, in fact, related) to the hierarchical growth of dark matter halos \citep{bcek91,lc93}. The barriers for both the models are higher at larger redshifts implying that relatively fewer random walks would cross the threshold, with the typical first crossing occurring at larger \So\ or smaller bubble size; in other words, fewer bubbles with smaller sizes will form at earlier times.
Other than sharing this generic feature related to hierarchical growth, the barriers for the ESP and FZH04 models are qualitatively quite different from each other. We first discuss the nature and consequences of these differences and then provide a physical justification of their origin.

First, at fixed redshift, we see that the ESP barrier is higher than the corresponding FZH04 one. As discussed above, this implies fewer bubbles with smaller sizes in the ESP model as compared to FZH04. 
One can thus expect that the total fractional volume contained within ionized regions $Q$ would be smaller 
in the ESP case as compared to FZH04. 

If the only difference between the ESP and FZH04 barriers were a constant vertical shift, then one would also expect that that the bubble size distribution shows a smaller characteristic radius in ESP as compared to FZH04. However, we see from Figure~\ref{fig:barriers} that the \emph{slope} of the ESP barrier at each redshift is also larger than the corresponding FZH04 one. An increase in barrier slope tends to decrease the typical first crossing \So\ by making it more difficult for walks to cross at large \So, and hence \emph{increases} the characteristic bubble size. This is most easily seen in \eqn{fc-FZH-lin} for the case of walks with uncorrelated steps in the presence of a linearly increasing barrier: increasing the slope $B_1$ shifts the exponential cutoff due to the term ${\rm e}^{-B_1^2\So/2}$ to smaller values of \So\ and hence larger bubble sizes. (It also decreases the normalisation due to the term ${\rm e}^{-B_0B_1}$, but this can be countered by an appropriate decrease of $B_0$.) A qualitatively similar change in characteristic size occurs for walks with correlated steps (see equation~\ref{fc-HII}).
This effect would therefore compete with the one due to the relative difference in barrier height in determining the shape of the bubble size distribution.

\begin{figure}
\centering
\includegraphics[width=0.45\textwidth]{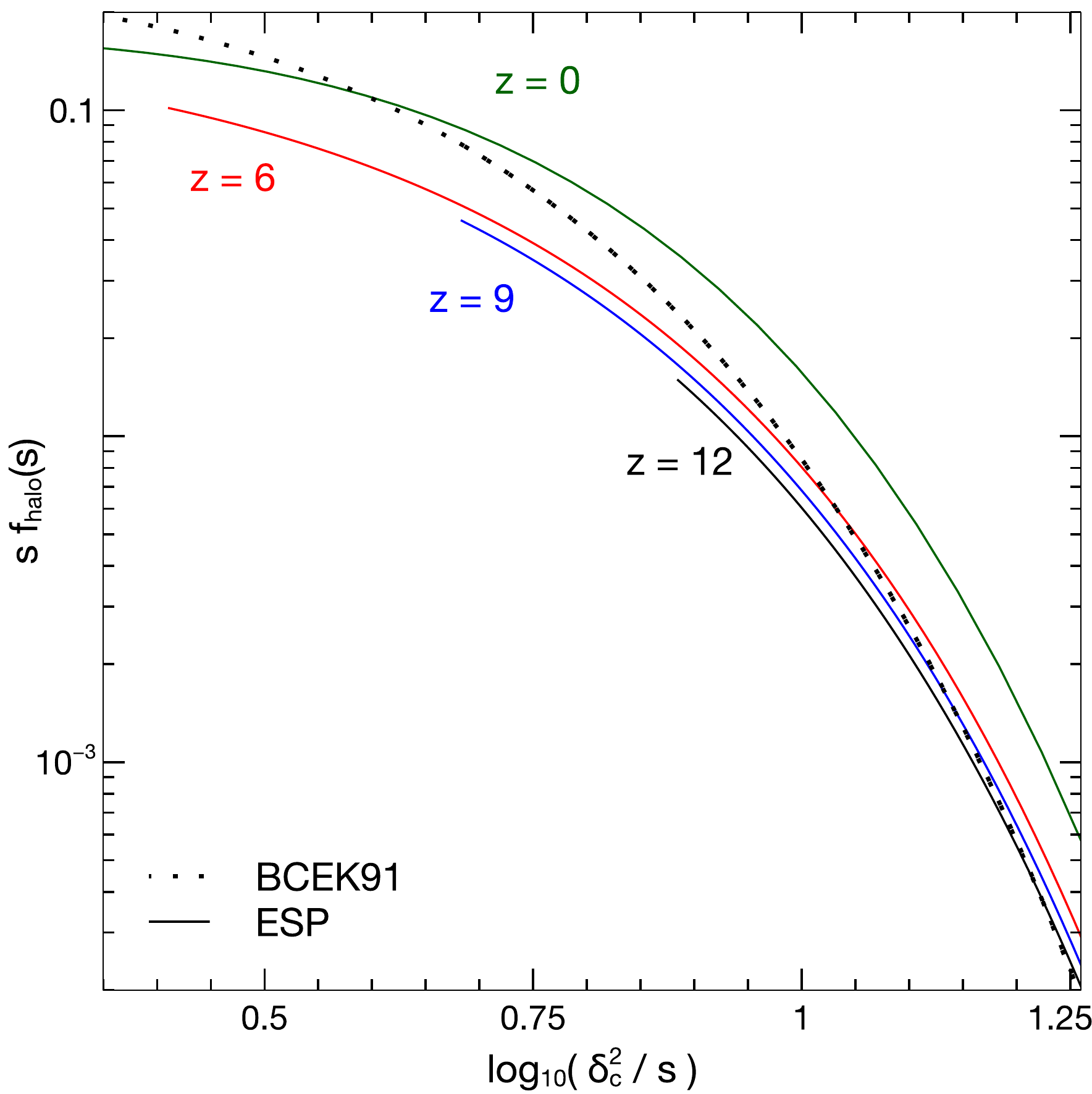}
\caption{The mass fraction $s f_{\rm halo}(s)$ in halos of mass $m$ where $s=\sig_0^2(m)$, as a function of $\delc^2(z)/s$ at different redshifts. The original excursion set prediction of \citet[][dotted black, marked BCEK91]{bcek91} is universal and therefore identical at all redshifts. The ESP prediction of \citet[][solid, from top to bottom $z=0,6,9,12$]{psd13} is non-universal and predicts lower mass fractions near the minimum mass at higher redshifts. This explains the difference in overall barrier heights of the FZH04 and ESP models as discussed in the text.}
\label{fig:sfsHalo}
\end{figure}

The reason for the difference in overall height of the ESP barriers as compared to FZH04 can be understood as follows. Recall that the collapse fraction $f_{\rm coll}(M,V)$ is an integral over the mass fraction $f(m|M,V)$ in halos of mass $m$ as discussed in Section~\ref{sec:formalism}. 
Since the mass function is steeply falling, the collapse fraction is dominated by halo masses close to the minimum mass $m_{\rm min}$ at any redshift. Figure~\ref{fig:sfsHalo} shows the unconditional halo mass fractions $sf_{\rm halo}(s)$ as a function of $\nu^2\equiv\delc^2(z)/s$ at different redshifts as predicted in the original excursion set calculation of \citet[][dotted black line]{bcek91} and in ESP (solid lines). The prediction for the former is universal ($sf_{\rm halo}(s)=\nu\,{\rm e}^{-\nu^2/2}/\sqrt{2\pi}$) and therefore independent of redshift in this plot, while in the latter case (equation~\ref{fESP} using $sf(s)=\nu f(\nu)/2$) the mass fraction decreases at higher redshifts (the curves from top to bottom show $z=0,6,9,12$). 
All the $z>0$ ESP curves start at\footnote{Most of the redshift dependence of $\nu_{\rm min}$ arises from $\delc(z)\propto D(0)/D(z)$ where $D(z)$ is the linear theory growth factor.} $\nu=\nu_{\rm min} = \delc(z)/\sig_0(m_{\rm min}(z))$.
The difference in the height of the barriers at a given redshift is therefore largely driven by the fact that at high redshifts the ESP calculation predicts fewer collapsed objects near the minimum mass than the original excursion set approach, making it more difficult to ionize the Universe.

The non-universality in the ESP halo mass fraction arises from the mass dependence of the quantities $V_\ast$ and $\gam$ that enter the calculation (see Appendix~\ref{app:esp}). This effect was discussed by \citet{psd13} and was shown to agree well with the non-universality in low redshift halo mass functions calibrated by \citet{t+08}. The ESP mass function has not yet been tested at high redshifts; in future work we will perform such a test to assess the level of accuracy of this predicted non-universality during the epoch of reionization. In section~\ref{subsec:sameQ} below we will study the case when we normalise both models to have the same global ionized volume fraction, in which case the overall height of the barriers becomes irrelevant.

The reason for the difference in \emph{slopes} of the ESP and FZH04 barriers, on the other hand, is related to the peaks constraint in ESP which is absent in the FZH04 model. 
To understand why, let us make some drastic approximations which still retain the qualitative features of the problem. (For technical details of the full calculation see Appendix~\ref{app:details}.) To start with, 
since we are interested in a fixed value of the collapse fraction (equal to $1/\zeta$), let us assume that we only need to study the single scale $s_\ast = s(m_\ast)$ where the characteristic mass $m_\ast$, as discussed above, depends on redshift and the values of \delo\ and \So. The overall relative behaviour of the two models then follows if we understand the behaviour in each model of the \emph{typical} walk in \delo\ as the smoothing scale \So\ changes, such that \delo\ approaches the halo barrier for $\So\to s_\ast$ and $\delo\to0$ for $\So\to0$. (For the purposes of this qualitative argument we can safely assume that the halo barrier is simply $\delc(z)$.) This typical walk is closely related to the \emph{density profile} surrounding the locations of patches in the initial density that are destined to form halos. (To see this, note that the profile gives the typical density as a function of distance from the chosen location, while the walk gives the density integrated in concentric spheres centered at this location.) Finally, the key point is that density profiles around peaks of \del\ are \emph{steeper} than those around randomly chosen locations. This is easy to understand on physical grounds, since in order for $\vec{x}$ to be a peak, the density in all directions around $\vec{x}$ must decrease by definition, while this is not required for an arbitrary point \citep[for a detailed calculation, see equations 7.8-7.11 and Figure~8 of][]{bbks86}. Consequently, the typical walk of interest to us (which is a proxy for the {\sc Hii} barrier in this case) is a steeper function of \So\ when constrained to be centered on peaks as compared to random locations\footnote{Essentially the same reasoning explains why the so-called peak-background split halo bias in the ESP model is higher at large masses than the corresponding quantity associated with, e.g. the \citet{st99} or \citet{t+08} mass functions, a point that was discussed by \citet{psd13}. These arguments rest on the excursion set ansatz which says that the number of halos in a given region at late times can be predicted by studying the environments of individual points (here, peaks) in the initial conditions. In the present case one is further identifying some of these environments (as determined by the {\sc Hii} barrier) as being ionized bubbles, which explains why the number of bubbles is sensitive to the peaks constraint even though the bubbles themselves are not necessarily centered on peaks.}.

\begin{figure}
\centering
\includegraphics[width=0.45\textwidth]{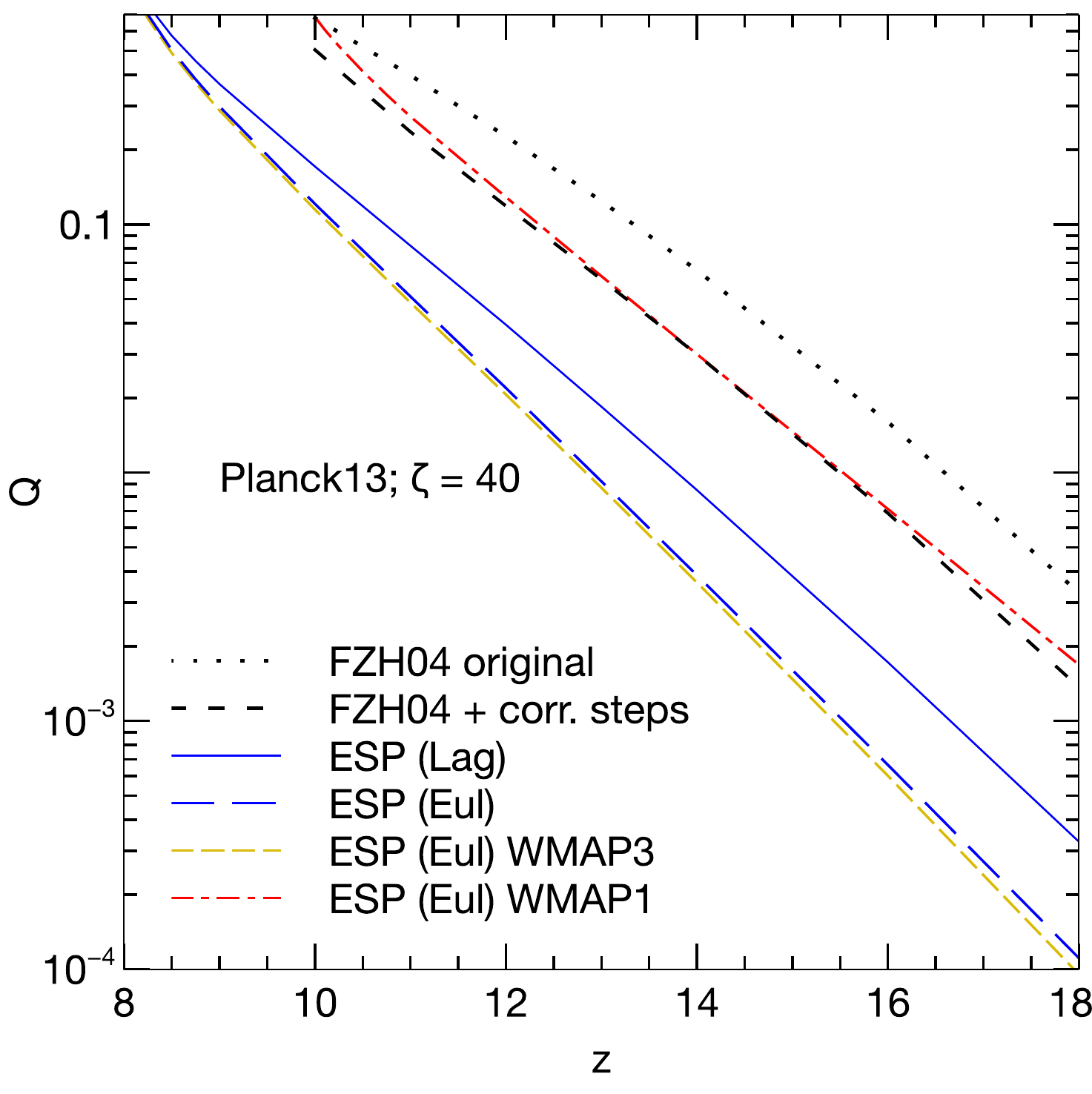}
\caption{The evolution of the volume fraction $Q$ in ionized regions for different models used in the paper. See text for details.}
\label{fig:Qvsz}
\end{figure}

Figure~\ref{fig:barriers} also shows linear approximations to the barriers for the two models. For the FZH04 case, the linear fit (dotted black) is the one in \eqn{B-FZH-lin}, which is the same as used by FZH04 and essentially corresponds to a straight line with a slope equal to the slope of the barrier at $\So = 0$. It is clear that the fit is quite good for small \So\ and is reasonable for higher values of \So. On the other hand, the linear approximation used for the ESP case (solid red) is actually a least-squares fit to the full calculation. As one can see, the fit is excellent for a wide range of \So\ values. We have checked that qualitatively similar results are obtained when using different values of the cosmological parameters or ionization efficiency. Hence from now on, we will work with the linear fits for the both FZH04 and ESP cases.

\subsubsection{Comparison of $Q(z)$}
\noindent
We next show in Figure \ref{fig:Qvsz} the evolution of $Q$ as a function of $z$ for different barriers. 
Unless stated otherwise, all curves in this plot use the Planck13 cosmology and $\zeta=40$.
The dotted black line shows the \emph{Lagrangian} ionized fraction from the original FZH04 calculation (equation~\ref{QLag-FZH}).
If we use  the same linear barrier $B_{\rm FZHlin}$ (equation~\ref{B-FZH-lin}) as FZH04 but the MS12 approximation with correlated steps for the bubble distribution, i.e., use the barrier \eqref{B-FZH-lin} in \eqn{fc-HII}, we obtain a Lagrangian $Q(z)$ as
shown by the dashed black curve (marked ``FZH04 + corr. steps''). The effect of correlated steps reduces the value of $Q$ by about a factor 2 at $z \gtrsim 12$.
The solid blue line shows the Lagrangian prediction for $Q(z)$ in the ESP case.
Clearly, the effect of including ESP is much more drastic than that of accounting for correlated steps alone as it reduces the value of $Q$ by nearly a factor of $10$ at $z=16$ and about a factor $5$ at $z=12$ compared to the FZH04 case. This was anticipated in our discussion of the barrier shapes above. 

\begin{figure}
\centering
\includegraphics[width=0.45\textwidth]{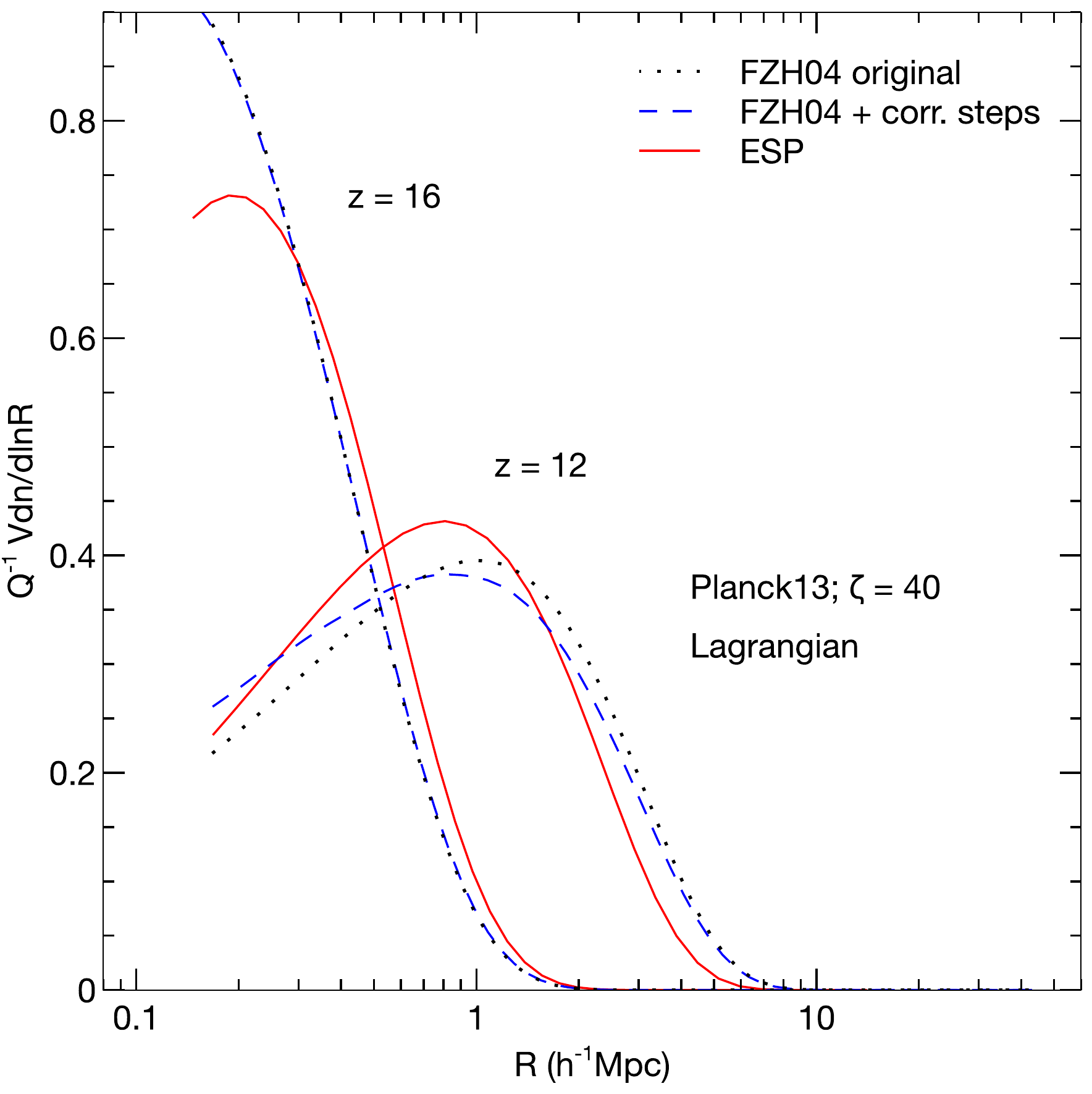}
\caption{Normalised Lagrangian size distributions of ionized bubbles for three cases: original FZH04 calculation (dotted black), FZH04 linear barrier with the MS12 approximation for correlated steps (dashed blue) and the ESP prediction (solid red). The curves are shown for two different redshifts with $\zeta=40$ for the Planck13 cosmology.}
\label{fig:dndlnR-ESPvsFZH}
\end{figure}

All the three cases discussed so far are for Lagrangian bubble sizes. When we convert to Eulerian sizes in the ESP model, the value of $Q$ (long-dashed blue curve) decreases further. This is a consequence of the fact that bubbles form preferentially in mildly overdense regions and hence their evolved sizes would be smaller than the corresponding Lagrangian ones. While the differences with the original FZH04 are now considerably larger at all redshifts, the difference with respect to the Lagrangian ESP calculation is not that dramatic (which was also anticipated by FZH04). Overall, then, the ESP Eulerian model at fixed $\zeta$ predicts a substantially different reionization history than the FZH04 model.
 
For later comparison, we also show $Q(z)$ for the ESP Eulerian case for the WMAP1 (dot-dashed red) and WMAP3 (fine-dashed yellow) cosmologies. 
The WMAP1 curve is substantially higher than the corresponding Planck13 curve, which is a consequence of the higher value of $\sig_8$ in WMAP1.
The fact that the Planck13 and WMAP3 curves are very close to each other means that changing the values of $\Omega_{\rm m}$ and $\Omega_{\rm b}$ has a much smaller effect on the reionization history, a feature we will see again when discussing the bubble size distributions below.

\subsubsection{Comparison of bubble size distributions}
\noindent
Our next goal is to see the effect of correlated steps and ESP on the bubble size distributions. We plot the normalized \emph{Lagrangian} size distribution $Q^{-1} V \der n/\der \ln R$ for different cases in Figure~\ref{fig:dndlnR-ESPvsFZH} for two redshifts $z = 12, 16$. As before, we compare results from the original FZH04 calculation (dotted black curves), the FZH04 linear barrier with correlated steps (dashed blue curves), and the Lagrangian ESP prediction (solid red curves). All curves use the Planck13 cosmology with $\zeta=40$.

One can see that the effect of correlated steps is noticeable at relatively lower redshifts and almost neligible at $z = 16$. At $z=12$, we see that the effect is to decrease the location of the maximum and lengthen the tail at lower radii.
The effect of including ESP in the formalism is more prominent. At $z=12$ it shifts the distribution to lower bubble sizes compared to the original FZH04, while this is reversed at $z=16$. The discussion in section~\ref{subsec:barriers} shows that the overall effect of ESP is the combination of two competing effects, namely a decrease in typical size due to the larger overall height of the ESP barrier and an increase due to its steeper slope as compared to FZH04.
Combined with the large relative difference in the predictions for $Q(z)$ seen in Figure~\ref{fig:Qvsz}, this shows that the number of bubbles $\der n/\der \ln R$ of a given radius $R$ would be very different for the two models.

\begin{figure*}
\centering
\includegraphics[width=0.9\textwidth]{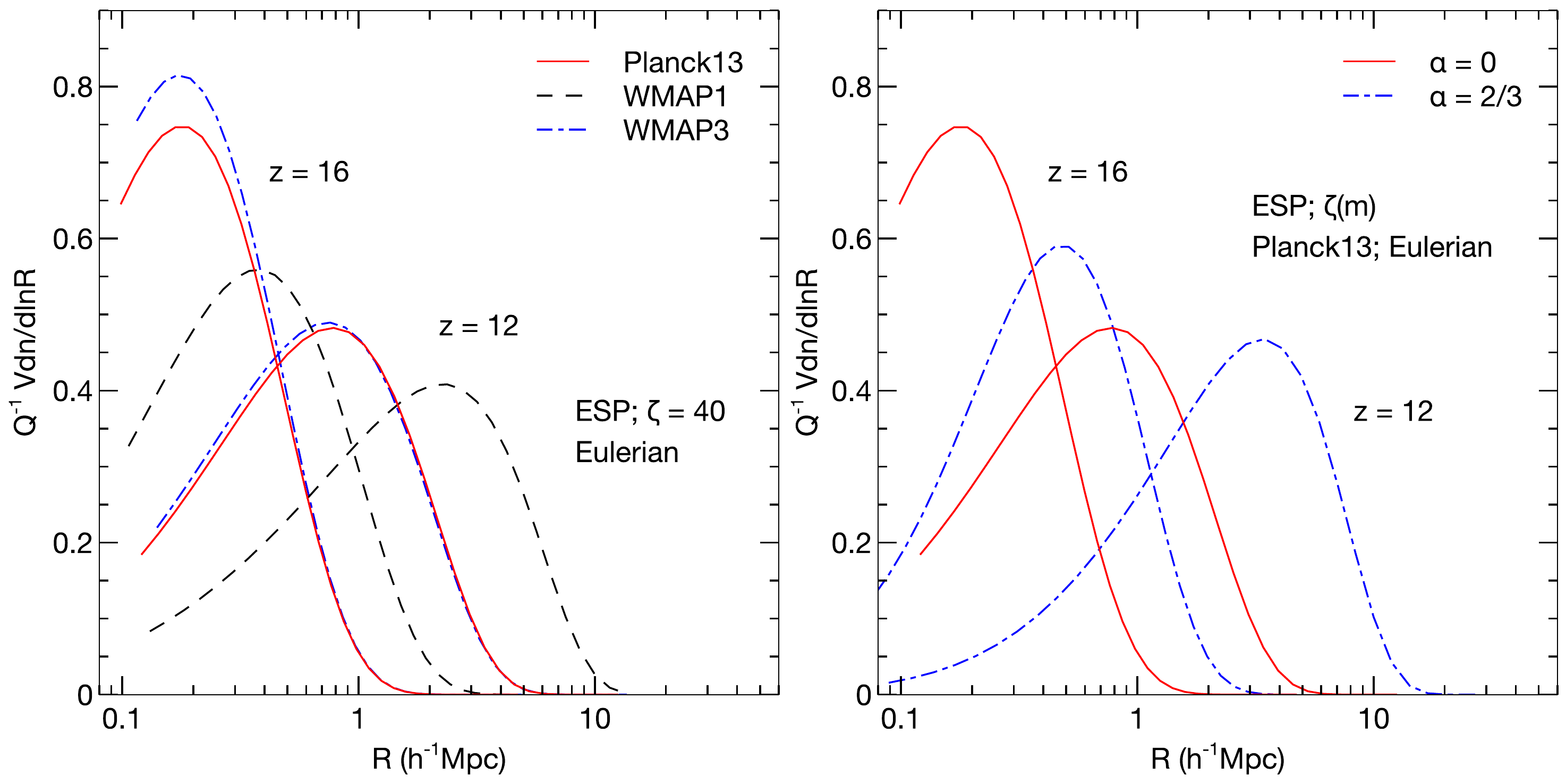}
\caption{
\emph{Left panel}: Normalised Eulerian bubble size distribution at two different redshifts with $\zeta=40$ for three cosmologies; Planck13 (solid red), WMAP1 (dashed black) and WMAP3 (dot-dashed blue). \emph{Right panel}: The same for the Planck13 cosmology but for $\zeta(m)\propto m^{\alpha}$ for two choices $\alpha=0$ (solid red) and $\alpha=2/3$ (dot-dashed blue). Note that we use the same efficiency for both models. The solid red curves at each redshift are identical across the two panels.
}
\label{fig:dndlnR-cosmozeta}
\end{figure*}

\subsubsection{Effect of cosmology and form of $\zeta$}
\label{subsec:cosmozeta}
\noindent
Having understood the main effects of including ESP in the formalism, we now proceed to see the effect of varying the cosmological parameters on the bubble distribution. We already saw the effect on $Q(z)$ in Figure~\ref{fig:Qvsz}. The left panel of Figure~\ref{fig:dndlnR-cosmozeta} shows the normalised \emph{Eulerian} bubble distributions for the Planck13 (solid red), WMAP1 (dashed black) and WMAP3 (dot-dashed blue) cosmologies, at $z=12,16$. These Eulerian distributions are very similar to the corresponding Lagrangian ones apart from a small shift towards smaller radii in the former (e.g., compare the solid red curves in the left panel of Figure~\ref{fig:dndlnR-cosmozeta} with the solid red curves in Figure~\ref{fig:dndlnR-ESPvsFZH}). The differences are non-negligible mainly at high redshift; this is because at high-$z$ the peak of the distribution is close to or smaller than $R_{\rm 0,min}$, so that a small shift in sizes leads to relatively large shifts in $Q$ and the shape of the distribution. At lower redshifts the characteristic bubble size is significantly larger than $R_{\rm 0,min}$ and small shifts have a much smaller impact. This also explains the relative difference between the solid and fine-dashed red curves of Figure~\ref{fig:Qvsz}.

Recall that the main difference between Planck13 and WMAP1 is in the value of $\sigma_8$ while the main difference between Planck13 and WMAP3 is in the value of $\Omega_{\rm m}$ and $\Omega_{\rm b}$. 
As with $Q(z)$ in Figure~\ref{fig:Qvsz}, the bubble distributions for Planck13 and WMAP3 are almost identical, indicating that the effect of varying the density parameters $\Omega_{\rm m}$ and $\Omega_{\rm b}$ is negligible. On the other hand, the effect of varying $\sigma_8$ is quite drastic as can be seen by comparing the curves for Planck13 and WMAP1. Increasing $\sigma_8$ leads to higher $Q$ and larger bubble sizes.

In our fiducial base model, we have taken $\zeta$ to be independent of the halo mass $m$. The parameter $\zeta$ is a combination of various astrophysical quantities like the initial mass function of stars within galaxies, the stellar spectra, the fraction of baryonic mass converted to stars and the escape fraction. The values and evolutionary properties of most of these quantities are unknown. Hence it is natural to investigate when $\zeta$ is allowed to be a function of $m$. 
This is straightforward in practice, since the only change required in the formalism is that the condition \eqref{fullyionized} which sets the barrier upon using \eqn{fcoll-ESP} must be replaced with the condition
\be
\int_{\So}^{s_{\rm min}}\der s\,f_{\rm ESP}(s|\delo,\So)\zeta(m(s)) \geq 1\,.
\notag
\ee
Since the integral over the conditional ESP mass fraction was numerical to begin with, this introduces no additional complication in our calculations.

We choose $\zeta$ to have a mass-dependent form $\zeta(m) = 40 \left(m/10^{8.1} h^{-1} M_{\odot}\right)^{\alpha}$.
We choose two values of $\alpha = 0$ (fiducial case) and $2/3$. The case $\alpha=2/3$ is motivated by observations of low-mass galaxies in the nearby universe \citep{2003MNRAS.341...54K} which suggest that the efficiency of the conversion of baryons into stars in galaxies with masses $< 3 \times 10^{10} M_{\odot}$ increases with halo mass, perhaps as a result of supernova feedback processes.. The comparison between the two cases is shown in the right hand panel of Figure \ref{fig:dndlnR-cosmozeta} (solid red for $\alpha=0$ and dot-dashed blue for $\alpha=2/3$; the solid red curves for each $z$ in the two panels of the Figure are identical). We can see that when we take the large mass haloes to be relatively more efficient than the small ones, the peak of the bubble size distribution shifts to larger radii. This is qualitatively consistent with the results of \citet{2006MNRAS.365..115F}.

\subsection{Models with the same $Q$}
\label{subsec:sameQ}
\noindent
So far we have compared different models keeping the value of $\zeta$ fixed. We found that the improvements in the excursion set formalism could significantly affect the height and shape of the ionization barrier, and consequently the global ionization fraction $Q$ as well as the bubble size distribution. However, one can argue that the value of $\zeta$ is highly uncertain, and hence any change arising from the mass functions in two models can possibly be compensated by changing the value of $\zeta$. In that case, one may never be able to distinguish between the two models
observationally (unless one has an independent constraint on $\zeta$, either theoretically or from observations). For example, \citet{2006MNRAS.365..115F} argued that this is indeed the case when one modifies the original FZH04
calculation to one using the \citet{st02} conditional mass function, particularly for larger values of $Q$. The main reason for this is that, although the collapsed mass fraction is different for the two mass functions, the bubble size distribution depends on the nature of the scale and density dependence of the conditional mass functions, which is quite similar for the cases analysed by \citet{2006MNRAS.365..115F}.

In order to see if this is also the case when comparing ESP and FZH04, we compare these two models at a given redshift $z = 8$ for the Planck13 cosmology keeping the value of $Q$ fixed. In practice, we do this by adjusting the value of $\zeta$ (we only consider constant $\zeta$ in this exercise) separately for both models, until each model leads to the desired value of $Q$. We choose two values $Q=0.25,0.75$ and only display results for the Lagrangian calculation; the Eulerian calculation would lead to qualitatively similar conclusions. 

\begin{figure*}
\centering
\includegraphics[width=0.9\textwidth]{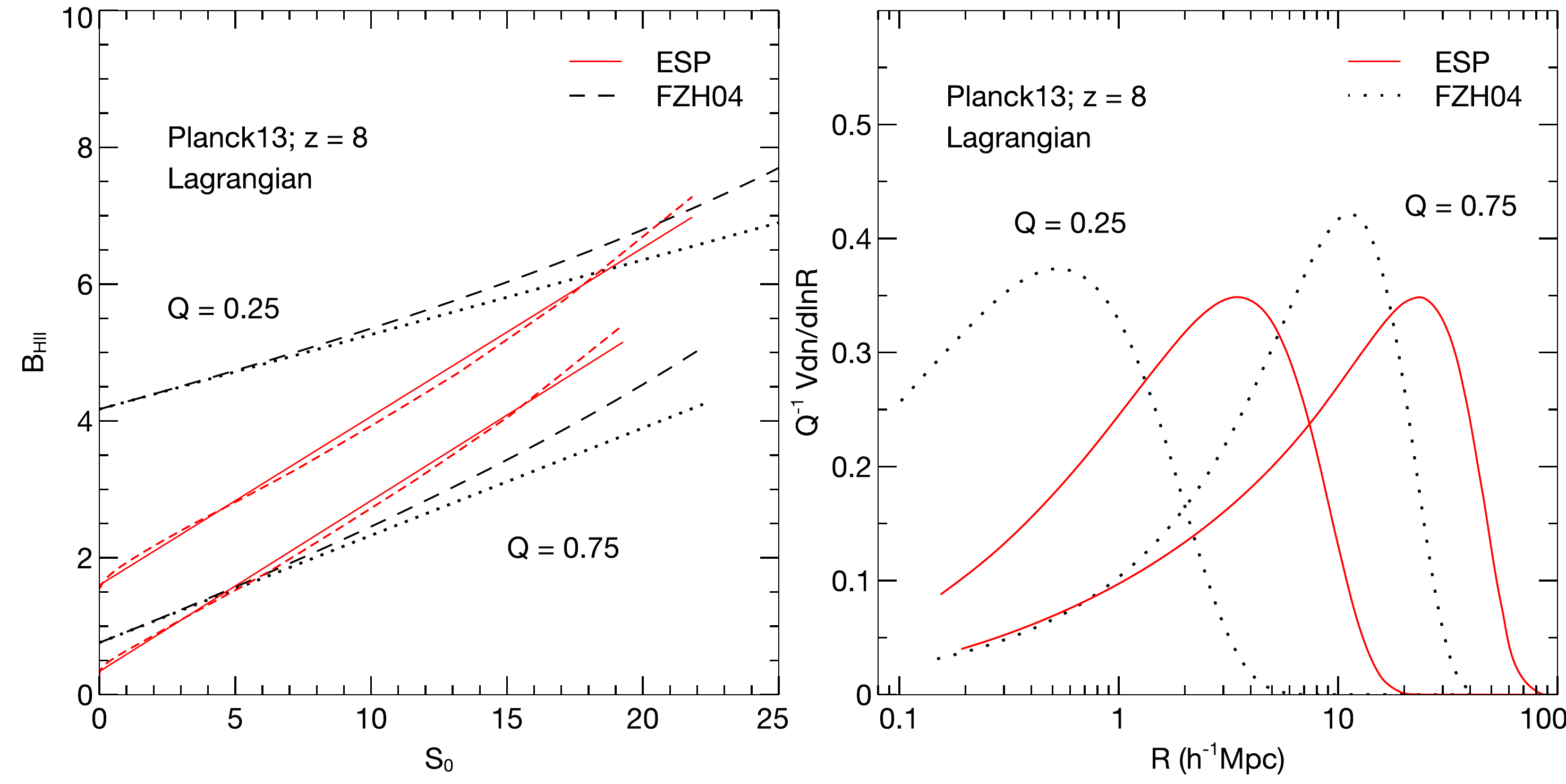}
\caption{Comparison of the FZH04 and ESP calculations at redshift $z=8$ for the Planck13 cosmology at the same value of global ionized Lagrangian volume fraction $Q$. We show comparisons at two fixed values $Q=0.25,0.75$. Note that the scales of the axes in this Figure are different from the previous Figures. \emph{Left panel}: Ionization barriers. The format of this panel is identical to Figure~\ref{fig:barriers}. The ESP barriers are systematically steeper than the FZH04 ones. \emph{Right panel}: Normalised Lagrangian bubble size distributions corresponding to the linear barriers of the left panel. The ESP calculation (solid red) predicts substantially larger bubble sizes than FZH04 (dotted black), even at $Q=0.75$.}
\label{fig:fixQ}
\end{figure*}

The ionization barriers are shown in the left hand panel of Figure~\ref{fig:fixQ}; the format of this panel is the same as in Figure~\ref{fig:barriers}. 
There are marked differences between the barriers for the two cases even when we normalize the models to the same $Q$. Although the differences are smaller for larger $Q$, which is consistent with the results of \citet{2006MNRAS.365..115F}, they are quite substantial even for $Q=0.75$. 
We see that while the normalisation to the same $Q$ as FZH04 has lowered the overall height of the ESP barriers, they continue to be \emph{steeper} than the corresponding FZH04 ones (compare Figure~\ref{fig:barriers}). As discussed earlier, this relative steepness is the physical consequence of the peaks constraint in ESP, which is absent in the FZH04 model and is a genuine difference between the two models that is evidently not degenerate with the physics of reionization as captured by a constant value of $\zeta$. (Of course, it might be possible to compensate for the nature of the peaks constraint by allowing a different \emph{mass dependence} of $\zeta$ in the FZH04 model compared to ESP; we do not explore this here.)

As pointed out in section~\ref{subsec:barriers}, a steeper barrier will favour larger bubble sizes, and we see from the right hand panel of Figure \ref{fig:fixQ} that this is indeed the case. This panel shows the Lagrangian bubble size distributions for the linear barriers in the left panel, with the solid red and dotted black curves showing the results for ESP and FZH04, respectively. We find that the bubble sizes are generally much larger in the ESP case. The differences between the two models are relatively less for larger $Q$; nevertheless, even at $Q = 0.75$ the difference in the 
value of $R$ at the maximum of the distribution is more than a factor of two. The FZH04 model predicts almost no bubbles with $R\gtrsim 40 h^{-1}$ Mpc for $Q = 0.75$, while the ESP calculation shows bubbles as large as $60 h^{-1}$ Mpc.
This can potentially have a significant impact on the {\sc Hi} fluctuation signal at these scales.

One way to verify these differences between the two models would be to compare with semi-numeric and radiative
transfer simulations.
In fact, there is now substantial evidence showing that the semi-numerical models agree with
the radiative transfer simulations quite well \citep[e.g.,][]{2007ApJ...654...12Z,2011MNRAS.414..727Z,2011MNRAS.411..955M}.
In addition, \citet{2007ApJ...669..663M} showed that the analytical calculations of FZH04 had a reasonable
match with their semi-numerical model, which gave rise to the conventional wisdom that the analytical 
results of FZH04 agree with simulations.
However, a careful look reveals that the semi-numerical model of \citet{2007ApJ...669..663M} is quite different
from the semi-numerical models which were used for comparing with radiative transfer simulations. 
For example, \citet{2007ApJ...654...12Z}, while identifying ionized cells using
spheres of varying size, flag only the central cell in a spherical region as ionized, whereas
\citet{2007ApJ...669..663M} flag the whole sphere. The 21cmFAST code presented in \citet{2011MNRAS.411..955M} too is
different for a number of reasons \citep[e.g., these authors use a sharp-$k$ filter for identifying ionized regions and flag only the central cell as ionized, while][use a real-space spherical TopHat filter and flag the full spherical region]{2007ApJ...669..663M}. 
To our knowledge, the only work which comes close to comparing the semi-numerical model of \citet{2007ApJ...669..663M} with radiative
transfer simulations is  \citet{2011MNRAS.414..727Z}. However, the density and halo fields in \citet{2011MNRAS.414..727Z} are obtained from $N$-body simulations, while they were derived using perturbation theory and the old excursion set formalism in \citet{2007ApJ...669..663M}. Hence, there does not exist any detailed comparison between the original semi-numeric model of \citet{2007ApJ...669..663M} and numerical simulations, indirectly implying that the analytical calculations of FZH04 have never been carefully compared with the radiative transfer simulations. We will make detailed comparisons of our analytical model with simulations in future work.

\section{Conclusions}
\label{sec:conclude}
\noindent
In this paper, we have use an improved calculation of the mass function of collapsed haloes and applied it to calculate the size distribution of ionized bubbles during the epoch of reionization. 
Our mass function calculation incorporates recent developments in the excursion set approach \citep{ms12,ps12,psd13} which allow us to implement the effect of correlations in the steps in the random walks of the smoothed density induced by the smoothing filter and, more importantly, account for the fact that haloes form preferentially around density peaks. In particular, we have used the Excursion Set Peaks (ESP) framework described by \citet{psd13}. 
Our calculation of the bubble size distribution follows previous work presented by \citet[][FZH04]{fzh04} with the modifications mentioned above.
We find that these modifications have a significant impact on the size distribution of bubbles which we summarize here.

\begin{itemize}
\item Including the effect of correlated steps in the random walks systematically lowers the ionized volume fraction (Figure~\ref{fig:Qvsz}) as well as the characteristic bubble size, and enhances the low radius tail of the size distribution (Figure~\ref{fig:dndlnR-ESPvsFZH}).
\item The peaks constraint alters the conditional halo mass function and leads to steeper excursion set barriers in our calculation as compared to FZH04 (Figure~\ref{fig:barriers} and left panel of Figure~\ref{fig:fixQ}). This is true regardless of whether we compare the models using the same astrophysics of reionization (fixed $\zeta$) or keeping the ionized volume fraction the same (fixed $Q$). (See section~\ref{subsec:barriers} for a detailed discussion of the relative differences in barrier shapes.)
\item At fixed $\zeta$, the barrier heights in ESP are systematically larger than in FZH04 (Figure~\ref{fig:barriers}), leading to a significantly \emph{smaller} ionized volume fraction (Figure~\ref{fig:Qvsz}). The relative typical bubble sizes are determined by the competition between the relative heights and slopes of the barriers, and show a more complex behaviour (Figure~\ref{fig:dndlnR-ESPvsFZH}). 
\item At fixed $Q$, however, the primary difference between ESP and FZH04 is in the steepness of the barriers mentioned above, and the ESP bubble size distribution in this case peaks at substantially \emph{larger} sizes than the FZH04 one (Figure~\ref{fig:fixQ}). Although the relative difference decreases with increasing values of $Q$, we see differences in characteristic radius of a factor of $2$ even at values as large as $Q=0.75$. This is primarily due to the nature of the peaks constraint in the conditional halo mass function which has been absent in all previous calculations, and is contrary to earlier claims based on the Sheth-Tormen mass function \citep{2006MNRAS.365..115F}.
\item We have also explored the effect of cosmological parameters on the bubble size distribution (Figure~\ref{fig:dndlnR-cosmozeta}). We find, as expected, that changing the value of $\sig_8$ by $10\%$ has a much more dramatic effect on the bubble sizes than changing $\Omega_{\rm m}$ or $\Omega_{\rm b}$ by the same amount. Changing the form of the ionization efficiency $\zeta$ by allowing for a mass dependent $\zeta(m)$ in the ESP calculation shows effects qualitatively similar to those seen previously by \citet{2006MNRAS.365..115F}.
\item Additionally, we have also explicitly accounted for the time evolution of bubble sizes (Eulerian as opposed to Lagrangian sizes) and we find non-negligible effects at high redshift on the ionized volume fraction (Figure~\ref{fig:Qvsz}) and bubble size distribution (compare Figure~\ref{fig:dndlnR-ESPvsFZH} with left panel of Figure~\ref{fig:dndlnR-cosmozeta}), with much smaller differences at lower redshift. (See section~\ref{subsec:cosmozeta} for a discussion.)
\end{itemize}

In recent times, most calculations of the {\sc Hii} bubble size distribution and the corresponding {\sc Hi} signal have been based on numerical simulations \citep{2006MNRAS.369.1625I,2006MNRAS.372..679M,2007ApJ...654...12Z,2007MNRAS.377.1043M,2007ApJ...671....1T,2007A&A...474..365S,2008ApJ...681..756S,2008ApJ...680..962L,2009A&A...495..389B} and semi-numerical methods \citep{2007ApJ...669..663M,2008MNRAS.386.1683G,2011MNRAS.411..955M}. The disadvantage with using full numerical simulations is that they are computationally expensive and are often not suitable for probing large regions of parameter space. This is not the case with the semi-numerical methods which are quite fast. These methods use the excursion set formalism to calculate the collapsed fraction and ionized regions within a numerical realization of the dark matter density field. It is interesting to note here that most existing calculations of the collapsed fraction are based on the formalism of \citet[][i.e., along the lines of FZH04]{bcek91}, or scaled to match the Sheth-Tormen mass function \citep[see, e.g.,][]{2011MNRAS.411..955M}. The results of these calculations would be modified if one used the ESP collapsed fraction (equation~\ref{fcoll-ESP}) instead, and we have shown (section~\ref{subsec:sameQ}) that the differences could be significant. Hence we believe that our calculations would have implications for semi-numeric models as well.

Our work can be extended in several directions. An obvious next step will be to compare our results with numerical and semi-numerical simulations of ionized bubbles, carrying out detailed $N$-body simulations and possibly exploring various semi-numeric techniques available in the literature. The basic excursion set language for describing {\sc Hii} regions has already been validated in comparisons with numerical simulations \citep{2007ApJ...654...12Z}.
It will be interesting to see whether the modifications in the analytical calculation that we have presented improve the agreement with numerical simulations, particularly since the original FZH04 analytical calculations have never been carefully compared with radiative transfer
simulations (see discussions at the end of section~\ref{subsec:sameQ}).
The formalism can also naturally account for additional effects such as, e.g., a stochasticity in either the collapse fraction \eqref{fcoll} as a function of density and scale or the reionization efficiency $\zeta$ \citep{2006MNRAS.365..115F}, and also more exotic effects such as those due to `warm' dark matter \citep{hp14,sitwell+14}. %
The main limitation of our method at this stage is that it fails to account for effects of local high density regions with large recombination rates. These regions would remain self-shielded for longer times and hence can possibly halt the progress of ionized regions. They may also have significant effects on the topology of the ionized regions.  Self-consistently accounting for this will require extensions to the formalism that are beyond the scope of the present work.
Finally, it will be particularly interesting to calculate the power spectrum of the {\sc Hi} signal which would be directly observable in future 21cm experiments. This will involve a calculation along the lines of the halo model \citep[see, e.g.,][]{cs02}. 
We will return to these issues in future work. 

\section*{Acknowledgments}
We thank an anonymous referee for insightful comments that helped improve the quality of the presentation.

\bibliography{mnrasmnemonic,IGM-ADS,espeaks}

\appendix
\section{}
\label{app:details}
Here we collect details of some of the calculations used in the main text. The following spectral integrals will be useful:
\begin{align}
\sig_{j{\rm G}}^2 &= \int\der\ln k\,\Del(k)\,k^{2j}{\rm e}^{-k^2R_{\rm G}(R_{\rm L})^2}\,,\notag\\
\sig_{{\rm 1m}}^2 &= \int\der\ln k\,\Del(k)\,k^{2}{\rm e}^{-k^2R_{\rm G}(R_{\rm L})^2/2}W(kR_{\rm L})\,,\notag\\
\Sc &= \int\der\ln k\,\Del(k)\,W(kR_{\rm L})W(kR_0)\,,\notag\\
\sig_{{\rm 1m}\times}^2 &= \int\der\ln k\,\Del(k)\,k^{2}{\rm e}^{-k^2R_{\rm G}(R_{\rm L})^2/2}W(kR_0)\,,
\label{sigma-j}
\end{align}
where the Fourier transform of the TopHat filter is given by $W(y)=(3/y^3)(\sin y- y\,\cos y)$.
The presence of Gaussian filtering ensures that these integrals are always finite. The Gaussian smoothing scale $R_{\rm G}(R_{\rm L})$ is related to the Lagrangian (TopHat) scale $R_{\rm L}$ by requiring $\avg{\del_{\rm G}|\del_{\rm TH}} = \del_{\rm TH}$, i.e. $\avg{\del_{\rm G}\del_{\rm TH}} = \avg{\del_{\rm TH}^2} = \sig_0^2(R_{\rm L})$, where the subscripts `G' and `TH' denote Gaussian and TopHat smoothing, respectively \citep{psd13}. In practice this gives $R_{\rm G}(R_{\rm L}) \approx 0.46 R_{\rm L}$ with a slow variation. The last two equations define cross-correlations which will appear in the calculation of the conditional mass fraction. As in the main text, we will also use $s=\sig_0^2(m)$ and $\So=\sig_0^2(M)$ where $m\propto R_{\rm L}^3$ and $M\propto R_0^3$.

\subsection{Excursion set peaks (ESP)}
\label{app:esp}
The ESP model prescribed by \citet{psd13} is based on the first crossing, by peak-centered random walks, of a ``square-root'' barrier 
\be
B=\delc(z)+\beta\sqrt{s} = \delc(z)+\beta\sig_0(m)\,,
\label{sqrt-barrier}
\ee
which is inspired by the ellipsoidal collapse model \citep{smt01} and 
where $\beta$ is a stochastic variable whose distribution is set by requiring the distribution of walk heights at \emph{fixed} mass $p(\del|m)$ to match measurements in $N$-body simulations presented by \citet{rktz09}. In particular, \citet{psd13} showed that setting $p(\beta)$ to be Lognormal with mean $0.5$ and variance $0.25$ leads to self-consistent results which are robust against small changes in the latter values.

The mass fraction in halos is given by equations (12) and (14) of \citet{psd13}:
\begin{align}
f_{\rm ESP}(\nu) &= \int\der\beta\, p(\beta)\,f_{\rm ESP}(\nu|\beta)\,,
\label{fESP}
\end{align}
where
\begin{align}
\nu\,f_{\rm ESP}(\nu|\beta) &= \frac{V}{V_\ast} \int_{\beta\gam}^\infty\der x\left(x/\gam-\beta\right)F(x)\,p(B/\sig_0,x)\notag\\
&= \frac{V}{V_\ast}\frac{{\rm e}^{-(\nu+\beta)^2/2}}{\sqrt{2\pi}}
\int_{\beta\gam}^\infty\der x\left(x/\gam-\beta\right)F(x)\notag\\
&\ph{\int\der x(x/\gam-\beta)}\times
p_{\rm G}(x-\beta\gam-\gam\nu;1-\gam^2)\,,
\label{fESP-beta}
\end{align}
where $\nu\equiv\delc(z)/\sig_0(m)$, $p_{\rm G}(y-\mu;\Sigma^2)$ is a Gaussian in the variable $y$ with mean $\mu$ and variance $\Sigma^2$, and $F(x)$ is given by
\begin{align}
F(x)&=\frac12\left(x^3-3x\right)\left\{\erf{x\sqrt{\frac52}}+\erf{x\sqrt{\frac58}}
\right\} \notag\\
&\ph{x^3-3x}
+ \sqrt{\frac2{5\pi}}\bigg[\left(\frac{31x^2}{4}+\frac85\right){\rm
    e}^{-5x^2/8} \notag\\
&\ph{\sqrt{x^3-3x+\frac2{5\pi}}[]}
+ \left(\frac{x^2}{2}-\frac85\right){\rm
    e}^{-5x^2/2}\bigg]\,,
\label{bbks-Fx}
\end{align}
\citep[equations~A14--A19 in][BBKS]{bbks86}. Also, \gam\ and $V_{\ast}$ are spectral quantities that define the distribution of peaks (similar to those defined in BBKS):
\be
\gam\equiv\sig_{1{\rm m}}^2/(\sig_0\sig_{\rm 2G}) \quad;\quad V_\ast\equiv(6\pi)^{3/2}(\sig_{\rm 1G}/\sig_{\rm 2G})^3\,.
\label{gam-Vst}
\ee
The spectral ratio $\gamma$ is related to the width of the matter power spectrum 
while $V_\ast$ is related to the typical inter-peak separation and can be thought of as a characteristic peak volume (BBKS). For power-law power spectra $P(k)\propto k^n$ with $-3<n<1$, one can prove that $\gamma$ is constant while $V_\ast\propto V$, which means that the ESP mass fraction $f_{\rm ESP}$ for this case is explicitly universal, being a function only of the scaling variable $\nu$. For CDM-like spectra, on the other hand, $\gamma$ and $(V_\ast/V)$ both show weak but non-trivial dependencies on smoothing scale and hence mass, which means that the resulting mass function is predicted to be weakly non-universal: $f_{\rm ESP}=f_{\rm ESP}(\nu(m,z);\gamma(m),V_\ast(m))$. \citet{psd13} showed that this non-universality agrees well with the low redshift measurements of halo mass functions in $N$-body simulations performed by \citet{t+08}. As discussed in the main text, the non-universality also has consequences for the size distribution of ionized bubbles at high redshift.

To calculate the conditional mass fraction $\fcoll(M,V)$ we will use the approximation discussed by \citet[][MS12]{ms12} and implemented by \citet{mps12}. Namely, in the first line of \eqn{fESP-beta} we replace $p(B/\sig_0,x)\to p(B/\sig_0,x|\delo)$ where $p(B/\sig_0,x|\delo)$ is a conditional bivariate Gaussian. Strictly speaking, this only ensures that walks which upcross at scale $s=\sig_0^2(m)$ had height \delo\ at scale $\So=\sig_0^2(M)$, while we actually require the stronger condition that these walks must have remained below the barrier $B$ at all scales $s^\prime < s$. However, \citet{mps12} showed by comparing to the full numerical first crossing solution that the simpler approximation mentioned above remains very accurate unless the height \delo\ becomes close to the barrier itself. Since we never encounter this situation (the ionization barrier always works out to be significantly below the halo barrier), making this assumption gives a self-consistent calculation in our case. Further support for this approximation comes from the fact that using this approximation to compute halo bias gives an accurate description of measurements in $N$-body simulations \citep{psc+13,psd13}.

The conditional mass fraction is then given by \citep[][their equation 22]{psd13}:
\begin{align}
2sf_{\rm ESP}(s|\delo,\So) &= \int\der\beta\,p(\beta)\,\nu f_{\rm ESP}(\nu|\beta,\delo),
\label{fESPcond}
\end{align}
where
\begin{align}
\nu f_{\rm ESP}(\nu|\beta,\delo) &= (V/V_\ast) p(B/\sqrt{s}|\delo) \notag\\
&\ph{V/V_\ast}\times
\int_{\beta\gam}^\infty \der x\,(x/\gam-\beta)F(x)p(x|B,\delo)\,,
\label{fESPcond-full}
\end{align}
and it is understood that $B$ is evaluated at scale $s$ while \delo\ is the density contrast at scale \So.
If we define
\begin{align}
\Cal{Q} &\equiv 1-\frac{\So}{s}\left(\frac{\Sc}{\So}\right)^2\quad;\quad
\epc \equiv \frac{s\,\sig_{1{\rm m}\times}^2}{\Sc\sig_{\rm 1m}^2}\,,
\label{Q-epc-Gam-def}
\end{align}
then we have
\begin{align}
p(B/\sqrt{s}|\delo) &= p_{\rm G}\left(B/\sqrt{s}-\avg{\mu|\delo};{\rm Var}(\mu|\delo)\right)\,,\notag\\
p(x|B,\delo) &= p_{\rm G}\left(x-\avg{x|B,\delo};{\rm Var}(x|B,\delo)\right)\,,
\label{condGaussians}
\end{align}
with
\begin{align}
\avg{\mu|\delo}&=(\delo/\sqrt{s})(\Sc/\So)\quad;\quad{\rm Var}(\mu|\delo) = \Cal{Q}\,,
\label{mean-var-mu}
\end{align}
and
\begin{align}
\avg{x|B,\delo} &= \gam\left[\frac{\left(B-\delo(\Sc/\So)\right)}{\Cal{Q}\sqrt{s}}(1-\epc) + \frac{B}{\sqrt{s}}\epc\right]\,,\notag\\
{\rm Var}(x|B,\delo) &= 1-\gam^2 - \frac{\gam^2(1-\Cal{Q})}{\Cal{Q}}(1-\epc)^2\,,
\label{mean-var-x}
\end{align}
where \gam\ was defined in \eqn{gam-Vst}. The collapse fraction $\fcoll(M,V)$ is then given by \eqn{fcoll-ESP}.


\label{lastpage}

\end{document}